\documentclass[twocolumn,english,prl]{revtex4}
\usepackage[T1]{fontenc}
\usepackage[latin9]{inputenc}
\usepackage{color}
\usepackage{graphicx}

\makeatletter
\@ifundefined{definecolor}
 {\usepackage{color}}{}
\@ifundefined{definecolor}
 {\@ifundefined{definecolor}
 {\usepackage{color}}{}
}{}
\@ifundefined{definecolor}
 {\@ifundefined{definecolor}
 {\@ifundefined{definecolor}
 {\usepackage{color}}{}
}{}
}{}
\@ifundefined{definecolor}
 {\@ifundefined{definecolor}
 {\@ifundefined{definecolor}
 {\@ifundefined{definecolor}
 {\usepackage{color}}{}
}{}
}{}
}{}
\@ifundefined{definecolor}
 {\@ifundefined{definecolor}
 {\@ifundefined{definecolor}
 {\@ifundefined{definecolor}
 {\@ifundefined{definecolor}
 {\usepackage{color}}{}
}{}
}{}
}{}
}{}
\@ifundefined{definecolor}
 {\@ifundefined{definecolor}
 {\@ifundefined{definecolor}
 {\@ifundefined{definecolor}
 {\@ifundefined{definecolor}
 {\@ifundefined{definecolor}
 {\usepackage{color}}{}
}{}
}{}
}{}
}{}
}{}

\makeatother

\usepackage{babel}

\makeatother

\usepackage{babel}

\makeatother

\usepackage{babel}

\makeatother

\usepackage{babel}

\makeatother

\usepackage{babel}

\makeatother

\usepackage{babel}

\begin{document}

\title{Quantum Correlations in Two-Particle Anderson Localization}

\author{Yoav Lahini$^{1}$, Yaron Bromberg$^{1}$, \textcolor{black}{Demetrios
N. Christodoulides$^{2}$}, and Yaron Silberberg$^{1}$}

\affiliation{$^{1}$Department of Physics of Complex Systems, Weizmann Institute
of Science, Rehovot, Israel.}

\affiliation{$^{2}$CREOL/College of Optics, University of Central Florida, Orlando,
Florida, USA}
\begin{abstract}
We predict the quantum correlations between non-interacting particles
evolving simultaneously in a disordered medium. While the particle
density follows the single-particle dynamics and exhibits Anderson
localization, the two-particle correlation develops unique features
that depend on the quantum statistics of the particles and their initial
separation. On short time scales, the localization of one particle
becomes dependent on whether the other particle is localized or not.
On long time scales, the localized particles show oscillatory correlations
within the localization length. These effects can be observed in Anderson
localization of non-classical light and ultra-cold atoms. 
\end{abstract}
\maketitle
More than fifty years ago, P.W. Anderson predicted that a single quantum
particle released in a disordered lattice can exhibit exponential
localization in space \cite{Anderson}, a phenomenon termed Anderson
Localization (AL). Since then there has been an ongoing effort to
observe the signature of AL experimentally, for example using light
\cite{chabanov}. Recently a novel approach enabled the direct observation
in space of AL for photons \cite{TLoc,Pertch,segev,Lahini} and ultra-cold
atoms \cite{chabe}. These experiments, reporting the exponential
localization of the particle density distribution, agree with the
predictions of the single particle model as long as no interactions
are involved.

The question arises, whether there are measurable phenomena that are
not described by the single particle model when considering the localization
of a few photons or atoms. It is known that when indistinguishable
particles propagate together, exchange terms can result in the formation
of correlations between their positions even in the absence of interactions.
This result, known as the Hanbury Brown-Twiss (HBT) effect \cite{HBT},
was studied theoretically and experimentally for particles propagating
in free space. It was observed that quantum statistics play a crucial
role in the formation of HBT correlations - bosons tend to bunch,
while fermions exhibit anti-bunching \cite{HBT,Aspect}. Yet, it is
not clear how HBT correlations evolve when the particles propagate
through disordered media and become localized. While correlations
were recently observed in the multiple-scattering of non-classical
light \cite{Beenakker,vanExter,lodahl}, experiments on AL focused
on the particle density distribution, and therefore could not observe
spatial correlations between the localized particles. Now, experimental
techniques that allow a direct observation of localization \cite{segev,Lahini,chabe}
can provide access to the study of HBT correlations between particles
exhibiting AL.

In this Letter we predict the correlations between \emph{two indistinguishable}
quantum particles evolving simultaneously in disordered lattices.
We consider non-interacting particles, and therefore the particle-density
follows the single particle dynamics: both particles undergo Anderson
localization. Nevertheless, we find that the two particles develop
non-trivial spatial correlations due to interferences of all the scattering
paths the two particles can take \emph{as a pair}. On short time scales,
the localization of one of the particles uniquely determines whether
the other particle will be localized or not. On longer time scales,
when both particles localize, the particles exhibit oscillatory correlations
within the localization length. Remarkably, these oscillatory correlations
survive multiple scattering even after very long evolution times.
In addition, we show that fermionic correlations can be reproduced
by entangled bosonic states.

\begin{figure}
\includegraphics[clip]{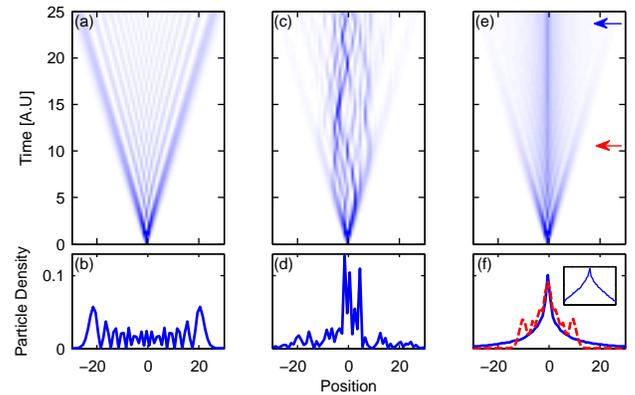}

\caption{\label{fig: 1}(color online). Dynamics of a single quantum particle
placed at $t=0$ on a single site. (a) In a periodic lattice, the
particle density distribution expands ballistically, with high density
at the edges of the distribution. (b) Cross-section of the density
distribution after some evolution. (c),(d) In a disordered lattice,
the expansion is limited to a finite region. (e) The density distribution,
averaged over 1000 realizations of disorder. The expansion starts
ballistically. After some propagation a localized component emerges
around the initial position, (red arrow, red cross-section in (f)).
The ballistic component decays in time, leaving the distribution exponentially
localized (blue arrow, blue cross-section in (f)). }

\end{figure}

To study two-particle localization, we follow the formalism developed
in \cite{Bromberg} for periodic systems. We consider a one dimensional
quantum tight-binding model, given by Hamiltonian:

\begin{equation}
H=\sum_{n}W_{n}a_{n}^{\dagger}a_{n}-\sum_{<n,m>}T_{n,m}a_{n}^{\dagger}a_{m}\label{eq:QH}\end{equation}
 where $a_{n}^{\dagger}$ is the creation operator for a particle
in site $n$, $W_{n}$ is the on-site energy and $T_{n,m}$ is the
tunneling amplitude between nearest neighbors. At time $t$ the creation
operator at site $r$ is given by \begin{equation}
a_{r}^{\dagger}(t)=\sum_{r'}U_{rr'}(t)a_{r'}^{\dagger}(t=0)\label{eq:adag}\end{equation}
 where the time-evolution operator $U_{rr'}(t)=(e^{i\hat{H}t})_{r,r'}$
is a unitary transformation given by calculating the exponent of the
Hamiltonian matrix $\hat{H}$, and describes the amplitude for transition
of a single particle located at site $r'$ at $t=0$, to site $r$
at time $t$. In the following, we study the evolution of two indistinguishable
particles in a disordered lattice, where disorder is introduced by
randomizing the tunneling amplitudes. We focus on the particle-density
$n_{r}(t)$=$\left\langle a_{r}^{\dagger}a_{r}\right\rangle $ and
on the two-particle correlation $\Gamma_{q,r}(t)$=$\left\langle a_{q}^{\dagger}a_{r}^{\dagger}a_{r}a_{q}\right\rangle $,
highlighting the single-particle versus two-particle features of the
dynamics.

\begin{figure}
\includegraphics[clip]{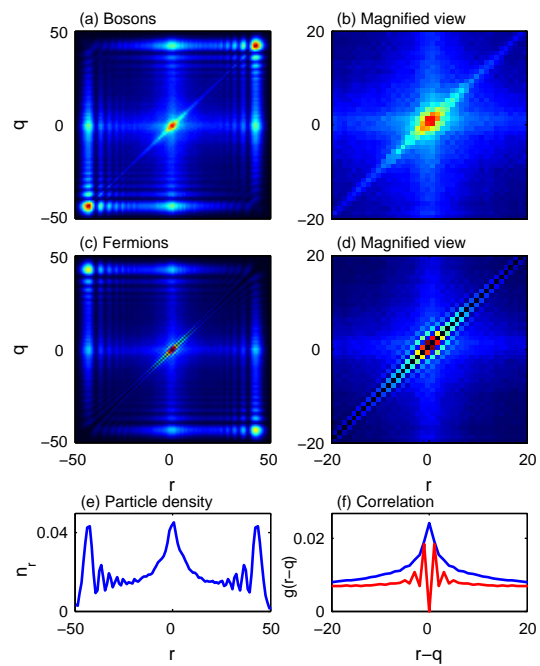}

\caption{\label{fig: 2}(color online). Two-particle correlation function in
disordered lattices. At $t=0$, two particles are placed at two neighboring
sites. The correlation function is calculated after a short evolution,
where remnants of the ballistic component are still present (see Fig.
1e,f), averaged over a 1000 disorder realizations. The matrices represent
the probability to detect one particle at site $q$ and one particle
at site $r$. (a) Bosons either scatter and become localized, or remain
ballistic. However, if they both remain ballistic, they always reach
the same side of the distribution. (b) Magnification of the localized
component at the center of the correlation map. (c) Fermions show
similar behavior, only that if they both remain ballistic they always
separate to different sides of the distribution. (d) Magnification
of the correlation map, showing a checker-like pattern. (e) The particle
density distribution, which is identical for fermions and for bosons,
computed by tracing out the position of one of the particles (summing
over the columns of the correlation matrix). (f) The inter-particle
distance probability for bosons (blue) and fermions (red). Bosons
tend to appear at the same site, while fermions are more likely to
be separated by an odd number of sites.}

\end{figure}

We start by considering the evolution of a single particle, described
by the particle density $n_{r}(t)=\left|U_{rr'}(t)\right|^{2}$ ,
where $r'$ is the location of the particle at $t=0$. In Fig. 1 we
plot the evolution of the particle density for an initial condition
$a_{0}^{\dagger}|0\rangle$, i.e. a single particle at site number
$0$. Fig. 1a depicts the propagation of the particle in a periodic
lattice, which is given by $n_{r}=J_{r-r'}^{2}(2Tt)$, where $J_{n}$
is the Bessel function of order $n$. This dynamics is known as a
quantum walk \cite{Bromberg,Qwalk,hagai,atomQwalk}, in which the
particle density expands ballistically (i.e. linearly in time), and
the probability to find the particle is highest at the edges of the
distribution (see Fig 1b). When disorder is introduced (here by randomizing
the tunneling amplitudes), the propagation changes significantly,
as shown in Fig. 1c,d. A statistical average of the probability distribution
over many realizations of disorder (Fig 1e,f) reveals a clear cross-over
from ballistic expansion at short times to exponential localization
at long times (blue arrow in Fig 1e). At intermediate times (red arrow
in Fig. 1e), the particle can be either localized or ballistic \cite{Lahini}.

We now consider two non-interacting particles which are initially
localized on sites $r'$ and $q'$, focusing on the correlations between
their locations at a later time $t$. The correlation matrix is given
by $\Gamma_{q,r}=\langle\langle|U_{qr'}U_{rq'}\pm U_{qq'}U{}_{rr'}|^{2}\rangle\rangle$
\cite{Bromberg}. Here $\langle\langle\cdot\rangle\rangle$ signifies
averaging over realizations of disorder, and the upper and lower signs
describe, respectively, bosons and fermions of same spin. The depicted
matrices (for example in Fig. \ref{fig: 2}) represent the probability
to detect at time $t$ exactly one particle at site $q$ and one particle
at site $r$. As we show below, the density distribution is identical
for bosons and for fermions (see Fig. \ref{fig: 2}e), yet the emerging
quantum correlations between the particle pair depends on the initial
position of each particle and on their quantum statistics.

We first consider the case in which initially the two particles are
positioned at two adjacent sites in a separable state, i.e. the initial
condition is given by $|\phi\rangle=a_{1}^{\dagger}a_{0}^{\dagger}|0\rangle$.
We calculate the two-particle correlation function after a relatively
short evolution time, in which each particle has a non-zero probability
to be localized or to remain ballistic (see Fig. 1). As shown by the
correlation matrix in Fig. 2a, if the two particles are bosons they
can both remain ballistic (corners of the correlation matrix), or
both become localized (center of the correlation matrix). The possibility
for one boson to localize and the other to remain ballistic is also
non-zero. However, when both bosons exhibit ballistic behavior, the
probability for the two bosons to separate to two different sides
of the distribution vanishes (top left and bottom right corners of
the correlation matrix). This is the signature of bosonic bunching
in lattices \cite{Bromberg}. In this case then, the two-particle
correlation has two components - a ballistic component showing spatial
bunching, and a localized component without spatial correlations.
A closeup on the correlation in the localized component is given in
Fig. \ref{fig: 2}b, also describing the correlation matrix at later
times, when the distribution is completely localized \cite{SUP}.

In the case of fermions, the correlation matrix of Fig. \ref{fig: 2}c,d
shows that if the two particles remain ballistic (i.e. both particles
did not localize yet), they will exhibit anti-bunching and separate
to opposite sides of the distribution. A closeup of the correlations
inside the localized component reveals a checkers-like pattern, meaning
that even when both fermions are localized they are non-trivially
arranged in space. Most significantly, this checker-like pattern survives
the random scattering process even after very long evolution times
\cite{SUP}.

\begin{figure}
\includegraphics[clip]{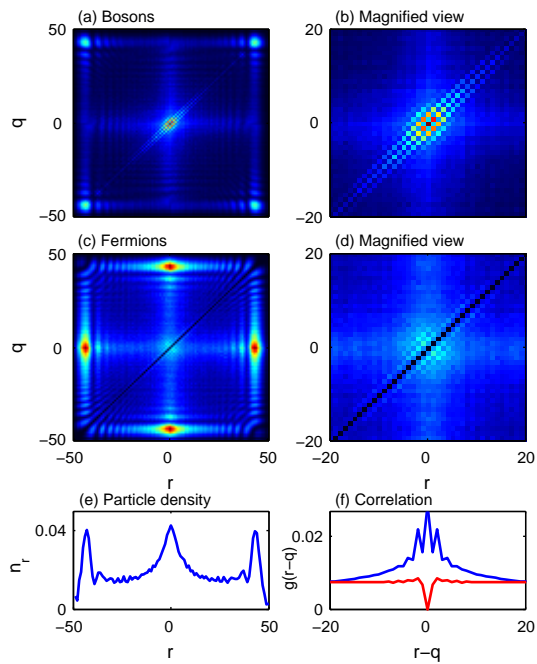}

\caption{\label{fig: 3}(color online). The two-particle correlation function
when At $t=0$, the two particles are placed at two non-adjacent sites.
(a) Bosons show two possible effects - they either both localized,
or both remain ballistic. (b) Closeup on the correlation inside the
localized component. (c) Fermions almost always split - one fermion
becomes localized and the other remains ballistic. (d) Closeup on
the correlation inside the localized component. (e) The particle density
distribution, identical for fermions and bosons. (f) The inter-particle
distance probability within the localized regime for bosons (blue
line) and fermions (red line). The fermions exhibit a flat inter-particle
distance distribution (except at zero separation), while the bosons
show an oscillating probability.}

\end{figure}

To understand the significance of the checker-like pattern we extracted
from the correlation matrix the inter-particle distance probability
given by $g(\Delta)=\sum_{q}\Gamma_{q,q+\Delta}$. Comparison between
the results for bosons and fermions are depicted in Fig. \ref{fig: 2}f.
Bosons (blue line) have the highest probability to be at the same
site, and the probability drops monotonously with distance. In contrast,
fermions have zero probability to be on the same site (as expected
due to Pauli exclusion). Yet, interestingly, the correlation oscillates:
the two fermions tend to be separated by an odd number of sites.

An even more dramatic effect takes place when the initial condition
is such that the two particles are placed at non-adjacent sites. Consider
the initial condition $|\phi\rangle=a_{-1}^{\dagger}a_{1}^{\dagger}|0\rangle$
- two particles at two sites separated by an empty site. Again we
calculate the correlation function after a relatively short propagation,
in which the dynamics includes both a ballistic and a localized component.
For bosons (Fig. \ref{fig: 3}a) there are only two possibilities:
either the particles are both localized (center of the correlation
matrix) \emph{OR} they both remain freely propagating (corners of
the correlation matrix). The possibility for one particle to be localized
and the other one to be free is diminished to nearly zero. A closeup
on the correlation when both particles are localized (Fig. \ref{fig: 3}b)
shows that in this case it is the bosons that exhibit the checkers-like
pattern; they are more likely to be separated by an \emph{even} number
of sites.

For this input state, fermions exhibit a different symmetry in the
correlations (Fig. \ref{fig: 3}c): the two particles tend to always
occupy different components of the expanding wavepacket: one fermion
becomes localized, while the other fermion remains freely propagating
in the ballistic component. The closeup on the localized regime (Fig.
\ref{fig: 3}d) shows a flat distribution, except for the zero probability
to be at the same site. The inter-particle distance for this case
is depicted in (Fig. \ref{fig: 3}f) \cite{SUP}.

\begin{figure}
\includegraphics[clip]{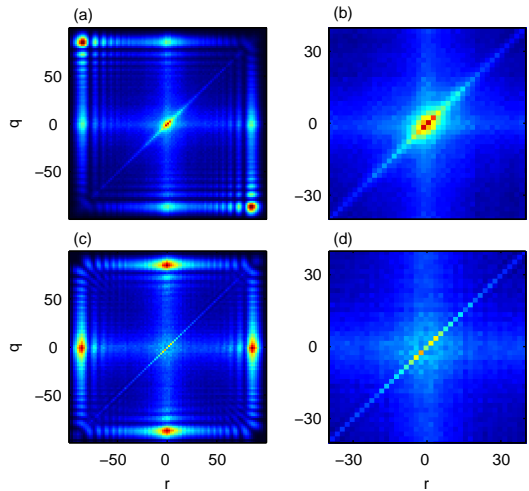}

\caption{\label{fig: 4}(color online) Correlation maps for path-entangled
bosons evolving in disordered lattices. (a) Correlation map for the
initial condition $|\phi\rangle=\frac{1}{2}[(a_{0}^{\dagger})^{2}+(a_{1}^{\dagger})^{2}]|0\rangle$.
(b) Magnified view of the localized component. (c) For $\phi=\frac{1}{2}[(a_{-1}^{\dagger})^{2}-(a_{1}^{\dagger})^{2}]|0\rangle$.
(d) Magnified view. These correlations are similar to the fermionic
results in Fig. 2c,d and 3c,d (see text). }

\end{figure}

Finally, it is of interest to consider the propagation in disordered
lattices of path-entangled input bosonic states, which can be regularly
generated in quantum-optics experiments. In a lattice system, such
a state is given by a superposition of two bosons - placed together
at one site \emph{OR} another site $|\phi\rangle=\frac{1}{2}[(a_{r'}^{\dagger})^{2}+e^{i\theta}(a_{q'}^{\dagger})^{2}]|0\rangle$.
Interestingly, properly constructed entangled states reproduce some
of the features presented by fermions. In Fig \ref{fig: 4} we present
the calculated correlation matrices for two kinds of path-entangled
input states: In Fig. \ref{fig: 4}a, we plot the correlation matrix
when the initial state is $|\phi\rangle=\frac{1}{2}[(a_{0}^{\dagger})^{2}+(a_{1}^{\dagger})^{2}]|0\rangle$.
The correlation matrix shows several features that are similar to
the ones exhibited by the fermions as seen in Fig. \ref{fig: 2}c
: the two particles can both become localized, both remain ballistic,
or split - one localized and one ballistic. However, they never end
up on the same ballistic lobe. The main difference is in the localized
region: the checker pattern disappears, and the diagonal $q=r$ of
the correlation matrix now shows enhanced probability. In Fig. \ref{fig: 4}c
we plot the correlation matrix when the initial state is $|\phi\rangle=\frac{1}{2}[(a_{-1}^{\dagger})^{2}+e^{i\pi}(a_{1}^{\dagger})^{2}]|0\rangle$.
i.e. the two states are separated by an empty site. This time the
pair shows a tendency to split - one boson is localized, while the
other remains ballistic. The correlation matrix is similar to that
exhibited by the fermion pair in Fig. \ref{fig: 3}c, except the enhanced
probability on the matrix diagonal $q=r$.

The emergence of the checker-like patterns in the correlations can
be explained in terms of the lattice eigenmodes excited in each case.
Disordered lattices support two kinds of eigenmodes: flat-phased and
staggered, in which adjacent sites are in phase /$\pi$ out of phase,
correspondingly \cite{Lahini}. Certain initial conditions, (depending
on the initial distance between the two particles and their quantum
statistics) involve the simultaneous excitations of both kinds of
eigenmodes\textcolor{black}{, resulting in a density pattern that
contains a component with two-site periodicity. This effect is washed
out in the density distribution averaged over all realizations of
disorder, as in each realization the oscillations appear in a different
location. However, the fact that such oscillations appears in each
realization will be recorded in the averaged correlation.}

At higher dimensional disordered systems, the single particle dynamics
can be qualitatively different. For example expanding wavepacket can
exhibit diffusive expansion before localization \cite{segev}. Nevertheless,
we have verified that the results described above appear also in two
dimensional lattices, including the checker-like patterns.

In conclusion, we studied the evolution of two non-interacting quantum
particles in disordered lattices exhibiting Anderson localization.
The two-particle correlation exhibits unique features that survive
the random scattering, such as oscillating correlations and correlated
cross-over from ballistic propagation to localization. The correlations
depend on the initial separation between the particles and on their
quantum statistics. Experimentally, the bosonic correlations can be
observed by injecting photon pairs generated using spontaneous parametric
down conversion, into disordered waveguide lattices. Such lattices
were recently used to observe Anderson localization of classical light
\cite{segev,Lahini}, and were shown to be feasible for observing
non-classical correlations between photon pairs \cite{hagai,Bromberg,Szameit}.
With atom-matter waves, correlations can be observed in the quantum
walks \cite{atomQwalk} of a pair of atoms in disordered lattices,
offering a possibility to study the effect of interaction \cite{Shepelyansky}
on correlations. Alternatively, Density-density correlations can be
measured for expanding Bose-Einstein condensates in disordered optical
potentials \cite{Altman,chabe}.

This work was supported by the German - Israel Foundation (GIF), the
Minerva Foundation and the Crown Photonics Center.

\end{document}